\documentstyle[amstex,amssymb,preprint,aps]{revtex}

\begin{document}
\title{The influence of single magnetic impurities on the conductance of quantum
microconstrictions.}
\author{A. Namiranian$^{\left( 1\right) }$, Yu.A. Kolesnichenko$^{\left(1,2\right) }$%
, A.N. Omelyanchouk$^{\left( 2\right) }$}
\address{$^{\left( 1\right) }$ {\small {\em ${}$ Institute for Advanced Studies in}}\\
Basic Sciences, \\
{\small {\em 45195-159, Gava Zang, Zanjan, Iran}} \\
$^{\left( 2\right) }${\small {\em \ B.Verkin Institute for Low Temperature}}%
\\
Physics and Engineering, \\
{\small {\em National Academy of Sciences of Ukraine,}} \\
{\small {\em 47 Lenin Ave., 310164 Kharkov, Ukraine}}}
\maketitle

\begin{abstract}
The nonlinear ballistic conductance of three-dimensional quantum
microconstrictions, which contain magnetic impurities, is investigated.
The nonlinear part of the conductance, which is due to the interaction of
electrons with magnetic impurities is obtained. The analytical results have
been analyzed numerically. It is shown that intensity of the Kondo anomaly
in the conductance as the function of the applied voltage depends on the
diameter of the constriction and the positions of impurities.
\end{abstract}

\bigskip The impurity-electron interaction in Kondo systems can be
effectively studied by using the point contacts (PC). In first measurements
of the differential PC resistance $R(V)$ in metals with magnetic impurities
the zero-bias Kondo anomaly had been observed \cite{Lysykh,Jancen,Naidyuk}.
These experiments were explained by quasiclassical theory of
Kondo effect in PC's \cite{OmTul}. It was shown that in second-order Born
approximation the magnetic impurity contribution to the PC resistance
includes the logarithmic dependence $R(V)\sim \ln \left( V\right) $ for $%
eV\gg T_{K}$ and the saturation for $eV\ll T_{K}$ ($T_{K}$ is the Kondo
temperature, $V$ is the voltage applied to the PC). In accordance to the
theory \cite{OmTul}, the nonlinear correction to the ballistic PC resistance
is proportional to the contact diameter. But in the experiments \cite{Lysykh,Jancen,Naidyuk}
the size dependence of the PC current was
not investigated due to the limited range of contact diameters, which were
accessible.

The development of the technique of mechanically controllable break
junctions (MCBJ) has made it possible to create the stable PC's, with the
diameter adjustable over broad range, down to a single atom \cite{Muller,Krans}.
In the MCBJ experiments \cite{Andrei,Yanson} authors
had studied the resistance of ultrasmall contacts with magnetic impurities
as function of the PC diameter $d.$ In the contrast to the prediction of the
quasiclassical theory \cite{OmTul} Yanson at el. \cite{Andrei,Yanson}
observed that Kondo scattering contribution to the contact resistance is
nearly independent from the contact diameter $d$ for small $d$. Such
behavior authors \cite{Andrei,Yanson} had explained by the increasing
of Kondo impurity scattering cross-section with decreasing of contact size.

In theoretical works \cite{KolOmTul} it was shown that in very small
contacts the discreteness of impurity positions must be taken into account,
and experiments \cite{Andrei,Yanson} may be explained by the
''classical'' mesoscopic effect of the dependence of the point contact
conductance on the spatial distribution of impurities. This effect is
essential in the ''short'' contacts and in the quasiclassical approximation
it disappears with the increasing of the contact length. Zarand and Udvardi
\cite{Zarand} had considered the contact in the in the form of a long
channel and suggested that the Kondo temperature $T_{K}$ is changed due to
the strong the local density of states fluctuations generated by the
reflections of conduction electrons at the surface of the contact. As a
result of that, the effective cross section of electrons has the maximum, if
the position of the impurity inside the contact corresponds to the maximum
in the local electron density of states. But the mesoscopic effect of the
spatial distribution of impurities in quantum contacts was not analyzed in
the paper \cite{Zarand}.

In ultrasmall contacts the quantum phenomena, which are known as quantum
size effect, occur. The effect of the $2e^{2}/h$ conductance quantization
has been observed in experiments on contacts in the two-dimensional electron
gas \cite{2Dfirst,2Dfirst-too} and in the ultrasmall
three-dimensional constrictions, which is created by using the scanning
tunnel microscopy \cite{STM1,STM2} and mechanically controllable
break junctions \cite{BJ1}. The defects produce the backscattering of
electrons, and thus break the quantization of the conductance. From the
other hand, the impurities situated inside the quantum microconstriction
produce the nonlinear dependence of the conductance on the applied voltage
\cite{interfer}. This dependence is the result of the interference of
electron waves reflected by these defects \cite{Agrait,Rutenbeek}. $%
~ $

In this paper we present the theoretical solution of this problem for the
conductance of a quantum microconstriction in the form of the long ballistic
channel , which contains single magnetic impurities. The study is made of
the first and second order corrections to the conductance of the ballistic
microconstriction in the Born approximation. The effect of impurity
positions is taken into account. Within the framework of the model of the
long channel the quantum formula for the conductance $G$ is obtained. By
using the model of the cylindrical microconstriction, the nonlinear
conductance as a function of voltage $V$ and the width of constriction $d$
is analyzed numerically for different positions of a single impurity.

Let us consider the quantum microconstriction in the form of a long and
perfectly clean channel with smooth boundaries and a diameter $d$ comparable
with the Fermi wave length $\lambda _{F}=h/\sqrt{2m\varepsilon _{F}},$ where
$\varepsilon _{F}$ is the Fermi energy. We assume that this channel is
smoothly (over Fermi length scale) connected with a bulk metal banks. As it
was shown \cite{Glazman,3Dtheor1}, in such constriction in the zeroth
approximation on the adiabatic parameter $\left| \nabla d\right| \ll 1$
accurate quantization can be obtained. The corrections to the tunneling and
reflection coefficients of electrons due to deviation from the adiabatic
constriction are exponentially small, except near the points where the modes
are switched on and off \cite{Yacobi}.

When a voltage $V$ is applied to the constriction, a net current start to
flow. In the limit $V\rightarrow 0,$ the ballistic conductance of the
quantum microconstriction is given by the formula
\begin{equation}
G=\frac{dJ}{dV}=G_{0}\sum_{\beta }f_{F}\left( \varepsilon _{\beta }\right)
,\quad
\end{equation}
where $f_{F}$ is the Fermi function, $\varepsilon _{\beta }$ is the minimal
energy of the transverse electron mode, $\beta $ is the full set of
transverse discrete quantum numbers. The ballistic quantum PC displays the
specific nonlinear properties, such as the conductance jumps $e^{2}/h.$ For
the two dimensional PC these effects was considered in the papers \cite{GKh,Zagoskin}.
 The aim of this study is to analyze the zero bias Kondo
minimum in the PC conductance. We assume that the bias $eV$ is much smaller
not only the Fermi energy $\varepsilon _{F}$, but also the distances between
the energies $\varepsilon _{\beta }$ of quantum modes. In this case the
effect of the influence of the applied bias to the transmission is
negligibly small.

Impurities and defects scatter the electrons that leads to the decreasing of
the transmission probability. In accordance with \ the standard procedure
\cite{KulYan,KOT} the decreasing of the electrical current $\Delta I$
due to the electron-impurity interaction connects with the velocity of the
energy $E$ dissipation by the relation:

\begin{equation}
\Delta IV=\frac{dE}{dt}=\frac{d\left\langle H_{1}\right\rangle }{dt};
\end{equation}
The Hamiltonian of the electrons $H$ contains the following terms:.

\begin{equation}
H=H_{0}+H_{1}+H_{int},
\end{equation}
where
\begin{equation}
H_{0}=%
\mathrel{\mathop{\sum }\limits_{k,\sigma }}%
\varepsilon _{k}c_{k\sigma }^{\dagger }c_{k\sigma }
\end{equation}
is Hamiltonian of free electrons,

\begin{equation}
H_{1}=\frac{eV}{2}%
\mathrel{\mathop{\sum }\limits_{k,\sigma }}%
sign(v_{z})c_{k\sigma }^{\dagger }c_{k\sigma }
\end{equation}
describes the interaction of electrons with electric field. The Hamiltonian
of the interaction of electrons with magnetic impurities $H_{int}$ can be
written as

\begin{equation}
H_{int}=\sum\limits_{j,k,k^{\prime }}{\bf J}_{j,k,k^{^{\prime }}}\left[
S_{z}\left( c_{k^{^{\prime }}\uparrow }^{\dagger }c_{k\uparrow
}-c_{k^{^{\prime }}\downarrow }^{\dagger }c_{k\downarrow }\right)
+S^{-}c_{k^{^{\prime }}\uparrow }^{\dagger }c_{k\downarrow
}++S^{+}c_{k^{^{\prime }}\downarrow }^{\dagger }c_{k\uparrow }\right] .
\end{equation}

Here the operator $c_{k\sigma }^{+}\left( c_{k\sigma }\right) $ creates
(annihilates) a conduction electron with spin $\sigma ,$ wave function $%
\varphi _{k},$ and energy $\varepsilon _{k};$ ${\bf S}${\bf \ }denotes the
spin of impurity; $v_{z}$ is the electron velocity along the channel; ${\bf J%
}_{j,k,k^{^{\prime }}}$ is the matrix element of the exchange interaction of
electron with impurity in the point ${\bf r}_{j}$; $k\sigma $ is the full
set of quantum numbers;

\begin{equation}
\;{\bf J}_{j,k,k^{\prime }}=\int d{\bf r\;}J({\bf r},{\bf r}_{j})\varphi
_{k}({\bf r})\varphi _{k^{\prime }}^{\ast }({\bf r}).
\end{equation}
The electron wave functions and eigenvalues in the long channel in the
adiabatic approximation are

\begin{align}
\varphi _{k}({\bf r})& =\psi _{\beta }({\bf R})\exp \left( \frac{i}{\hbar }%
p_{z}z\right) ; \\
\varepsilon _{k}& =\varepsilon _{\beta }+\frac{p_{z}^{2}}{2m_{e}};
\end{align}
where $k=\left( \beta ,p_{z}\right) ,$ $\beta $ is the set of discrete
transverse quantum numbers; $p_{z}$ is the momentum of an electron along the
contact axis; $m_{e}\ $is an electron mass; ${\bf r=}\left( {\bf R,}z\right)
,$ ${\bf R}$ is a coordinate in the plain, perpendicular to the $z$ axis.

Differentiating $\left\langle H_{1}\right\rangle $ over the time $t$ we
obtain the equation for the changing $\Delta I$ of the current as a result
of the interaction of electrons with magnetic impurities:

\begin{equation}
V\Delta I=\frac{1}{i\hbar }\left\langle \left[ H_{1}\left( t\right)
,H_{int}\left( t\right) \right] \right\rangle ,
\end{equation}
where
\begin{equation}
\left\langle ...\right\rangle =Tr\left( \rho \left( t\right) ...\right) .
\end{equation}
All operators are in the representation of interaction.

The statistical operator $\rho \left( t\right) $ satisfies to equation

\begin{equation}
i\hbar \frac{\partial \rho }{\partial t}=\left[ H_{int}\left( t\right) ,\rho
\left( t\right) \right] ,
\end{equation}
which can be solved using the perturbation theory:
\begin{equation}
\;\rho \left( t\right) =\rho _{0}+\frac{1}{i\hbar }\int\limits_{-\infty
}^{t}dt^{\prime }\left[ H_{int}\left( t^{\prime }\right) ,\rho _{0}\right] +%
\frac{1}{\left( i\hbar \right) ^{2}}\int\limits_{-\infty }^{t}dt^{\prime
}\int\limits_{-\infty }^{t^{\prime }}dt^{\prime \prime }\left[ H_{int}\left(
t^{\prime }\right) ,[H_{int}\left( t^{\prime \prime }\right) ,\rho _{0}]%
\right] \cdot \cdot \cdot
\end{equation}

By means of Eq.(13) the changing in the electric current due to magnetic
impurities can be determined

\begin{gather}
\Delta I=I_{1}+I_{2}+...= \\
-\frac{1}{\hbar ^{2}V}\int\limits_{-\infty }^{t}dt^{\prime }Tr\left( \rho
_{0}\left[ \left[ H_{1},H_{int}(t)\right] ,H_{int}(t^{\prime })\right]
\right) -  \nonumber \\
-\frac{1}{i\hbar ^{3}V}\int\limits_{-\infty }^{t^{\prime }}dt^{\prime \prime
}\int\limits_{-\infty }^{t}dt^{\prime }Tr\left( \rho _{0}\left[ \left[ \left[
H_{1},H_{int}(t)\right] ,H_{int}(t^{\prime })\right] ,H_{int}(t^{\prime
\prime })\right] \right) +...  \nonumber
\end{gather}

After the simple, but cumbersome calculations we find the first and second
order corrections to the PC current

\begin{equation}
I_{1}=-\frac{e\pi }{\hbar }s(s+1)\sum\limits_{n,m}\sum\limits_{i,j}\;({\it %
sign\ }v_{z_{m}}-{\it sign\ }v_{z_{n}})(f_{m}-f_{n})\ \ \delta \left(
\varepsilon _{n}-\varepsilon _{m}\right) \;{\bf J}_{j,n,m}\;{\bf J}_{i,m,n};
\end{equation}

\begin{gather}
I_{2}=\frac{\pi e}{\hbar }s(s+1)\sum\limits_{n,m,k}\sum\limits_{i,j,l}({\it %
sign\ }v_{z_{k}}-{\it sign\ }v_{z_{n}})\;\;\; \\
\lbrack \delta (\varepsilon _{n}-\varepsilon _{k})\Pr \;\frac{1}{\varepsilon
_{m}-\varepsilon _{k}}+\delta (\varepsilon _{m}-\varepsilon _{k})\Pr \;\frac{%
1}{\varepsilon _{n}-\varepsilon _{k}}]\;\;\;  \nonumber \\
\;[{\bf J}_{j,n,k}{\bf J}_{i,m,n}{\bf J}_{l,k,m}+{\bf J}_{j,k,n}{\bf J}%
_{i,n,m}{\bf J}_{l,m,k}]\;\;  \nonumber \\
\lbrack 2f_{n}(f_{k}-f_{m})+(f_{m}-f_{k})],\;\;\;  \nonumber
\end{gather}
where $f_{n}=f_{F}\left( \varepsilon _{n}+\frac{eV}{2}signv_{z}\right) .$
The first addition $I_{1}$ to the PC current$~$describes~the small
spin-depended correction (of the order $(J/\varepsilon _{F})^{2}$ ) to the
changing of the current due to the usual scattering. The second addition $%
I_{2}$ is also small too, but contains the Kondo logarithmic dependence on
the voltage, and it is most important for the analysis of the nonlinear
conductance of constrictions with magnetic impurities.

The expressions (15) and (16) can be further simplified in the case of $%
\delta -$potential of impurities

\begin{equation}
J\left( {\bf r}\right) =J\delta \left( {\bf r}\right)
\end{equation}
In this case the addition $I_{2}$ to the ballistic current has the form:

\begin{gather}
I_{2}=\frac{2J^{3}\pi e}{\hbar }s(s+1)\sum\limits_{n,m,k}\sum\limits_{i,j,l}(%
{\it sign\ }v_{z_{k}}-{\it sign\ }v_{z_{n}})\;\;\; \\
\lbrack \delta (\varepsilon _{n}-\varepsilon _{k})\Pr \;\frac{1}{\varepsilon
_{m}-\varepsilon _{k}}+\delta (\varepsilon _{m}-\varepsilon _{k})\Pr \;\frac{%
1}{\varepsilon _{n}-\varepsilon _{k}}]\;\;  \nonumber \\
\mathop{\rm Re}%
[\varphi _{k}^{\ast }({\bf r}_{j})\varphi _{n}^{\ast }({\bf r}_{i})\varphi
_{m}^{\ast }({\bf r}_{l})\varphi _{k}({\bf r}_{l})\varphi _{m}({\bf r}%
_{i})\varphi _{n}({\bf r}_{j})]  \nonumber \\
\lbrack 2f_{n}(f_{k}-f_{m})+(f_{m}-f_{k})],\;\;\;  \nonumber
\end{gather}

As it follows from the Eqs.(16), (18), the current $I_{2}$ depends from the
positions of impurities. Two effects influence by value $I_{2}:$ the effect
of quantum interference of scattered electron waves, which depends from the
distances between impurities, and effect of the electron density of states
in the points, where the impurities are situated. The nonlinear part of the
conductance can be easy obtained after differentiation the Eq.18 over the
voltage $G_{2}=dI_{2}/dV.$ In the case of a single impurity and at zero
temperature $T=0$ this equation can be analytically integrated over momentum
$p_{z}$ and the conductance $G_{2}$ takes the following form:
\begin{eqnarray}
G_{2} &=&-\frac{\pi e^{2}m_{e}^{3}}{\hbar ^{4}}J^{3}s(s+1)\sum_{\alpha
,\beta ,\gamma }\sum_{\varkappa =\pm }\left| \psi _{\alpha }({\bf R})\right|
^{2}\left| \psi _{\beta }({\bf R})\right| ^{2}\left| \psi _{\beta }({\bf R}%
)\right| ^{2}\left[ p_{\alpha }^{\left( \varkappa \right) }p_{\beta
}^{\left( \varkappa \right) }p_{\gamma }^{\left( \varkappa \right) }\right]
^{-1}\cdot \\
&&\left[ \ln \left| \frac{p_{\gamma }^{\left( \varkappa \right) }-p_{\gamma
}^{\left( -\varkappa \right) }}{p_{\gamma }^{\left( \varkappa \right)
}+p_{\gamma }^{\left( -\varkappa \right) }}\left( \frac{p_{\alpha }^{\left(
\varkappa \right) }}{p_{\gamma }^{\left( \varkappa \right) }}\right) \right|
+\left( 1-\delta _{\alpha \beta }\right) \ln \left| \frac{p_{\alpha
}^{\left( \varkappa \right) }p_{\beta }^{\left( -\varkappa \right)
}-p_{\alpha }^{\left( -\varkappa \right) }p_{\beta }^{\left( \varkappa
\right) }}{p_{\alpha }^{\left( \varkappa \right) }p_{\beta }^{\left(
-\varkappa \right) }+p_{\alpha }^{\left( -\varkappa \right) }p_{\beta
}^{\left( \varkappa \right) }}\right| \right. +  \nonumber \\
&&\left. \delta _{\alpha \beta }\ln \left| \frac{\left( p_{\alpha }^{\left(
\varkappa \right) }\right) ^{2}-\left( p_{\alpha }^{\left( -\varkappa
\right) }\right) ^{2}}{\left( p_{\alpha }^{\left( -\varkappa \right)
}\right) ^{2}}\right| \right] ;  \nonumber
\end{eqnarray}
where
\begin{equation}
p_{\alpha }^{\left( \pm \right) }=\sqrt{2m_{e}\left( \varepsilon _{F}\pm
\frac{eV}{2}-\varepsilon _{\alpha }\right) },
\end{equation}
and the transverse parts of the wavefunction $\psi _{\alpha }({\bf R})$ and
the electron energy $\varepsilon _{\alpha }$ are defined by Eqs. (8), (9).

$~~~$Carrying out the numerical calculations we use the free electron model
of a point contact consisting of two infinite half-spaces connected by a
long ballistic cylinder of a radius $R$ and a length $L$ (Fig.1). In a limit
$L\rightarrow \infty $ the electron wave functions $\varphi _{k}\left( {\bf r%
}\right) $ and eigenstates $\varepsilon _{k}$ can be written as

\begin{equation}
\varphi _{k}\left( {\bf r}\right) =\frac{1}{\sqrt{\Omega }J_{m+1}\left(
\gamma _{mn}\right) }J_{m}\left( \gamma _{mn}\frac{\rho }{R}\right) \exp
\left( im\varphi +\frac{i}{\hbar }p_{z}z\right) ;
\end{equation}

\begin{equation}
\varepsilon _{k}=\varepsilon _{mn}+\frac{p_{z}^{2}}{2m_{e}};\quad
\varepsilon _{mn}=\frac{\hbar ^{2}}{2m_{e}R^{2}}\gamma _{mn}^{2}
\end{equation}
and cylindrical coordinates ${\bf r=}\left( \rho ,\varphi ,z\right) $ with
the axis $z$ along the channel axis have been used. Here $k=(n,m,p_{z})$ are
the quantum numbers, $\Omega =\pi R^{2}L$ is the volume of the channel, $%
\gamma _{mn}$ are the n-th zero of the Bessel function $J_{m}.$ Because the
degeneration of the electron energy on azimuthal quantum number $m$ ( as a
result of the symmetry of the model), quantum modes with $\pm m$ give the
same contribution to the conductance. In this model the ballistic
conductance (1) has not only steps $G_{0},$ but also steps $2G_{0}$ \cite
{3Dtheor1,3Dtheor2}.

\bigskip In Fig.2 the dependence of the nonlinear conductance on the applied
bias is shown for the different positions of a single magnetic impurity
inside the channel. The results obtained confirm that the nonlinear effect
is strongly depend on the position of impurity. If the impurity is situated
near the surface of the constriction ${\bf r=R}$, where the square module of
the electron wave function is small, its influence to the conductivity is
negligible. This conclusion is confirmed by the calculations of the
dependence $G_{2}$ on the position of the impurity for different number of
quantum modes (Fig.3). Results indicate that the mesoscopic effect of the
impurity position is more essential for ultrasmall contacts, which contain
only few conducting modes, and $G_{2}$ has a maximum. The similar results is
obtained for the dependence of $G_{2}$ on the radius $R$ of the constriction
(Figs.4,5). In the single-mode constriction (Fig.4) the conductance $G_{2}$
displays much more stronger dependence on $R,$ than in the contact with five
conducting modes (Fig.5).

Thus, we have shown that in the long quantum microconstrictions the spatial
distribution of magnetic impurities influences to the nonlinear dependence
of the conductance on the applied voltage. This mesoscopic effect is due to
the strong dependence the amplitude of an electron scattering on the
positions of impurities. As a result of the reflection from the boundaries
of the constriction the electron wave functions, which correspond to the
finite electron motion in the transverse to the contact axis direction, are
the standing waves. If the impurity is situated near the point, in which the
electron wave function is equal to zero (near the surface of the
constriction or, for quantum modes with numbers $n>1,$ in some points
inside), its scattering of electrons is small. The fact the amplitude of the
Kondo minimum of the conductance of the quantum contact display the
mesoscopic effect of the dependence on the positions of single impurities.
This effect is most important in the case, when only few quantum modes are
responsible on the conductivity of the constriction.\newpage

{\bf Figure captions.}\quad\

Fig. 1. Schematic representation of a ballistic microconstriction in the
form of a long channel, adiabatically connected to large metallic
reservoirs. Magnetic impurities inside the constriction are shown.

Fig. 2. The voltage dependence of the nonlinear part of conductance $G_{2}$
(19) from the distance of the impurity from the contact axis ($2\pi
R=5.2\lambda _{F};$ $T=0;$ 1 - $2\pi \rho =1.5\lambda _{F}$ ; 2 -$2\pi \rho
=2.5\lambda _{F}$ ; 3 -$2\pi \rho =3.0\lambda _{F}$ ; 4 - $2\pi \rho
=3.5\lambda _{F}$ ).

Fig. 3. The dependence of $G_{2}$ (19) on the position of the impurity for
the different quantum modes in the constriction ($V=0.02\varepsilon _{F};$ $%
T=0;$ 1 - one mode ($2\pi R=3\lambda _{F}$); 2 - three modes($2\pi
R=4\lambda _{F}$); 3 - five modes ($2\pi R=5.3\lambda _{F}$); 4 - six modes (%
$2\pi R=6\lambda _{F}$)).

Fig. 4. The dependence of $G_{2}$ (19) on the radius of the constriction for
the single mode channel and different positions of the impurity ($%
V=0.02\varepsilon _{F};T=0;$1 - $2\pi \rho =0.5\lambda _{F}$ ; 2 -$2\pi \rho
=1.0\lambda _{F}$ ; 3 -$2\pi \rho =1.5\lambda _{F}$ ; 4 - $2\pi \rho
=2.0\lambda _{F}$ )

Fig. 5. The dependence of $G_{2}$ on the radius for the microconstriction
with five quantum modes and different positions of the impurity ($%
V=0.02\varepsilon _{F};T=0;$1 - $2\pi \rho =0.5\lambda _{F}$ ; 2 -$2\pi \rho
=1.5\lambda _{F}$ ; 3 -$2\pi \rho =2.5\lambda _{F}$ ; 4 - $2\pi \rho
=4.5\lambda _{F}$ )

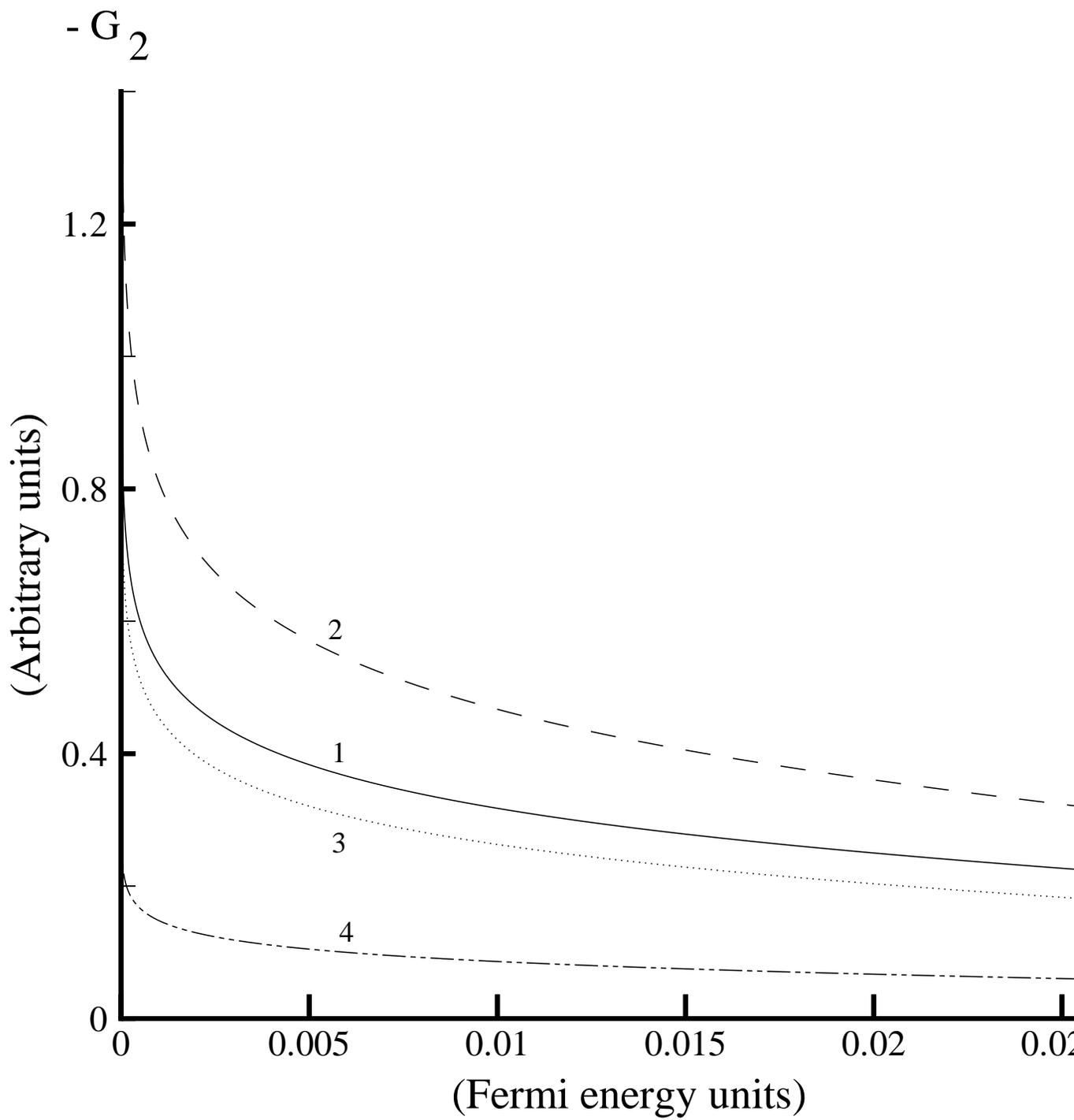
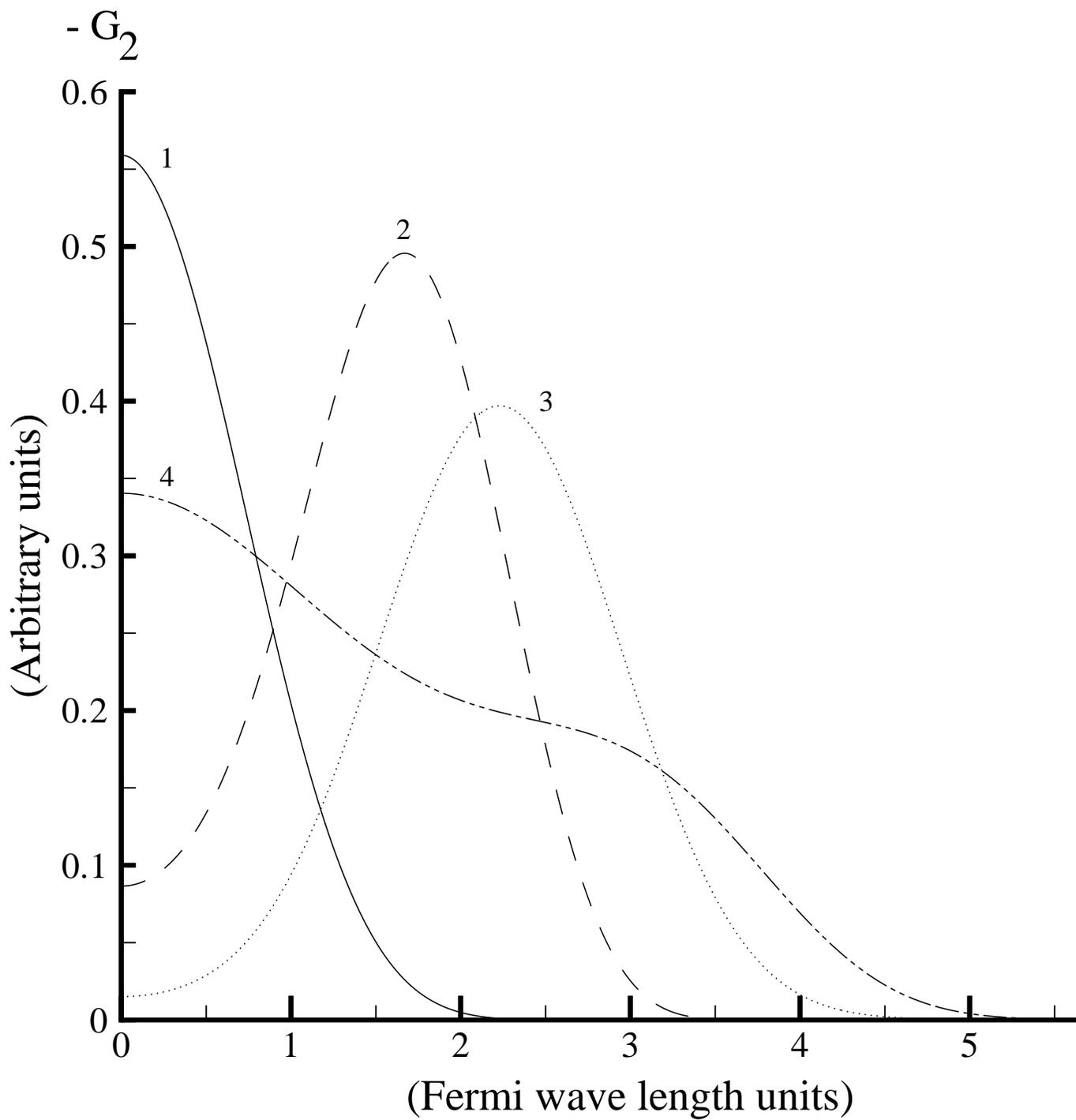
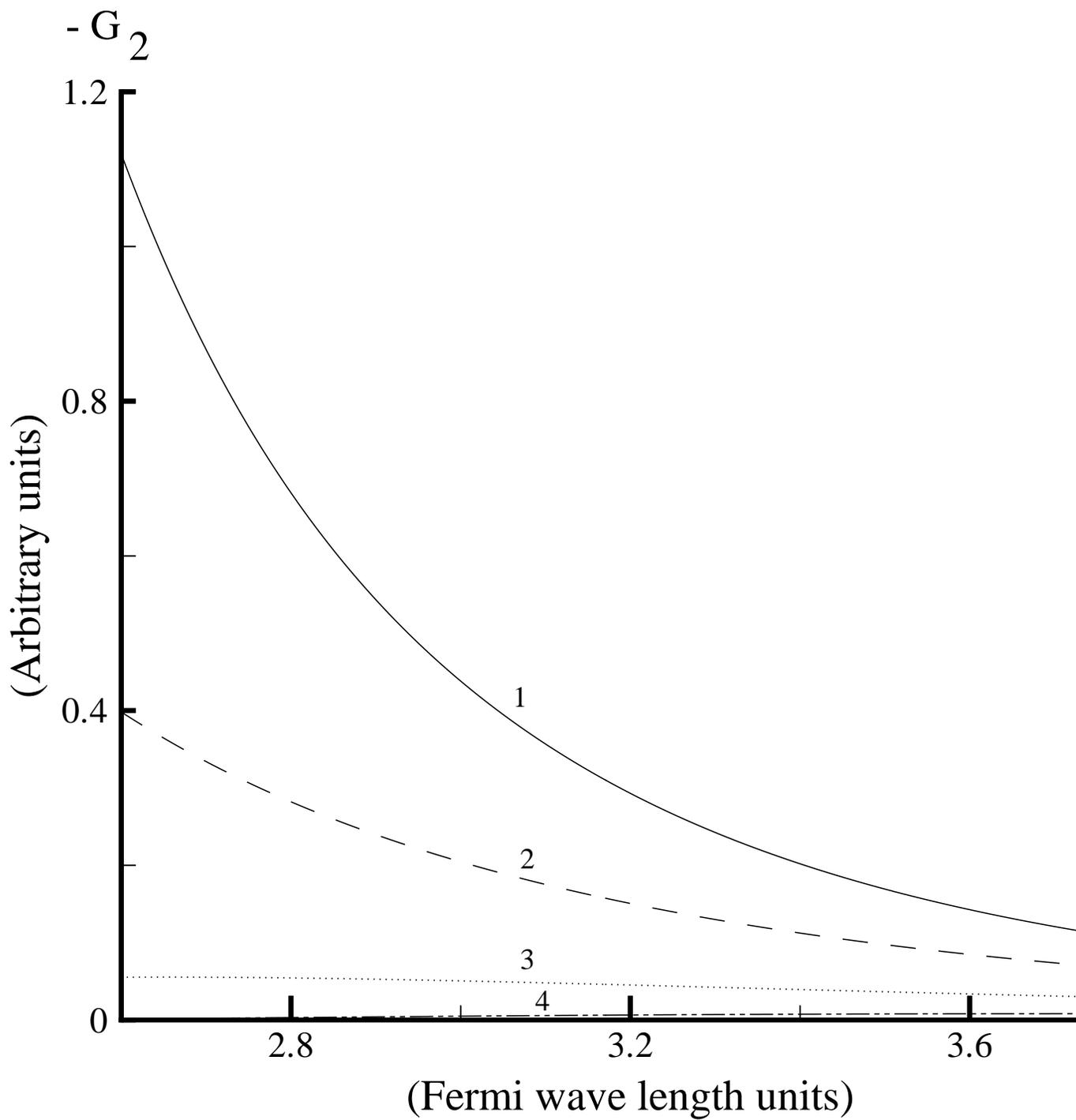
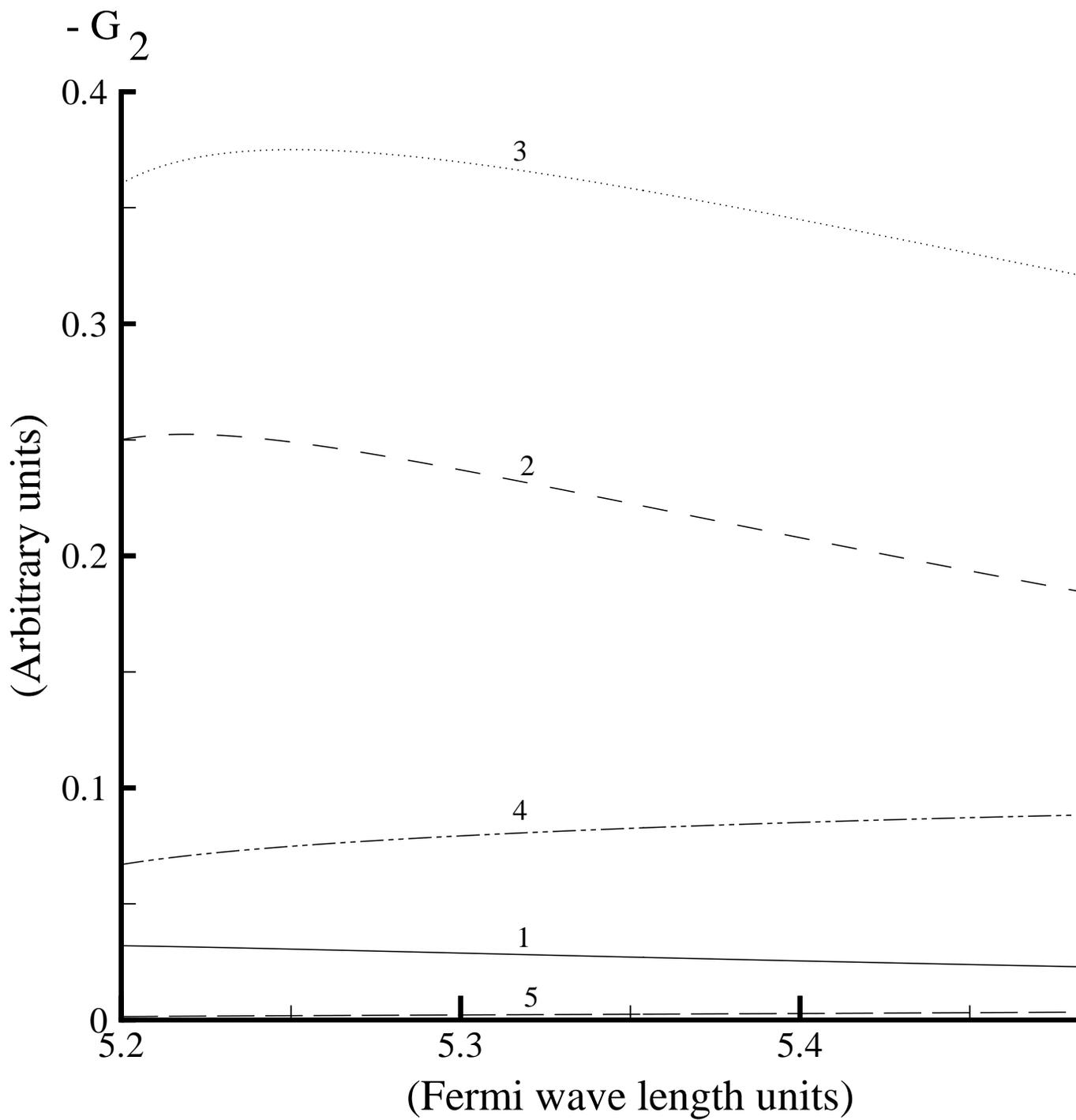
\end{document}